# Background-free time-resolved coherent Raman spectroscopy (CSRS and CARS): heterodyne detection of low-energy vibrations and identification of excited-state contributions

Pavel Kolesnichenko,[1,2] Jonathan O. Tollerud,[1] and Jeffrey A. Davis[1,2,a]

[1]*Centre for Quantum and Optical Science, Swinburne University of Technology, Melbourne, Victoria 3122, Australia*

[2]*Centre for Future Low-Energy Electronics Technologies, Swinburne University of Technology, Melbourne, Victoria 3122, Australia*

Coherent Raman scattering (CRS) spectroscopy techniques have been widely developed and optimized for different applications in biomedicine and fundamental science. The most utilized CRS technique has been coherent anti-Stokes Raman scattering (CARS), and more recently, stimulated Raman scattering (SRS). Coherent Stokes Raman scattering (CSRS) has been largely ignored mainly because it is often strongly affected by fluorescence, particularly for resonance enhanced measurements. However, in the cases of resonant excitation the information contained in the CSRS signal can be different and complementary to that of CARS. Here we combine the approaches of pulse shaping, interferometric heterodyne detection, 8-step phase cycling and Fourier-transform of time-domain measurements, developed in CARS and 2D electronic spectroscopy communities, to measure resonant CSRS and CARS spectra using a Titanium:sapphire oscillator. The signal is essentially background-free (both fluorescent and non-resonant background signals are suppressed) with high spectral resolution and high sensitivity, and can access low-energy modes down to ~30 cm$^{-1}$. We demonstrate the ability to easily select between CSRS and CARS schemes and show an example in which acquisition of both CSRS and CARS spectra allows vibrational modes on the excited electronic state to be distinguished from those on the ground electronic state.

## Introduction

Understanding the nature of low-energy (<200 cm$^{-1}$) vibrational modes is of high significance in several disciplines and can allow insights into the systems under investigation. In biological sciences, for example, low-frequency vibrations may be associated with large-scale nuclear motions in proteins, which can be associated with proteins' ability to tune their reactivity or enhance energy transfer in photosynthetic light-harvesting.[1,2] In condensed matter physics, low-frequency vibrations can, for example, characterize details of polycrystalline structures;[3,4] determine the size of nanomaterials (e.g. 1D single-wall carbon nanotubes,[5] 2D materials[6]); and may assist with identification of a low-temperature noncentrosymmetric phase of matter, such as those that host exotic type-II Weyl fermions.[7]

Low-frequency vibrations are typically measured by far-infrared (FIR)/terahertz (THz) spectroscopy,[8,9] or by various Raman scattering spectroscopy (RSS) techniques.[3–7,10,11] FIR/THz spectroscopy directly probes optically active low-energy vibrations, however, working in this spectral range poses several challenges: broad and strong water absorption can swamp the system response; generating and detecting light in this spectral range is difficult and lies in what is known as the THz gap. On the other hand, Raman techniques typically utilize visible, near-infrared (NIR), or near-ultraviolet (NUV) radiation, allowing the use of standard optical components and greater capabilities for measuring water containing samples. However, acquiring Raman spectra in low-energy spectral region usually requires sending the signal through triple-grating spectrographs[3,10] that have extremely low throughput and detection sensitivity, or by using expensive notch filters, based on ultrathin coatings and volume holographic gratings, with fixed laser wavelengths.[4–7,11]

Our approach to access low-energy vibrational modes is based on several ideas that have been separately implemented in coherent anti-Stokes Raman scattering (CARS) techniques[12] and multidimensional electronic spectroscopy (MDES)[13,14] (see Experiment). CARS is part of the family of coherent Raman spectroscopy (CRS) techniques that also includes coherent Stokes Raman scattering (CSRS)[15] and stimulated Raman scattering (SRS)[16,17] approaches. There are many similarities between CARS and CSRS schemes: both take advantage of coherent four-wave mixing processes that generate a signal only in specific directions defined by the phase matching condition, which drastically increases the detection sensitivity. Both CARS and CSRS spectra, however, suffer from the presence of non-resonant background (NRB). This background interferes with the signal of interest, and complicates the analysis[18], although with proper quantification the NRB response can be deconvolved[19] and in some circumstances can even enhance sensitivity.[18,20,21] In many cases, however, the noise from the NRB signal can reduce the sensitivity and resolution of the resonant Raman response, and hence eliminating the NRB is often desirable.[18] The spectral distortions coming from the NRB are even more pronounced in the low-energy region,[22–24] since most of the third-order NRB quantum pathways increase in efficiency with decreasing frequency. SRS methods were initially suggested to be intrinsically free from NRB, and while they are substantially better than CARS in this regard, there are still ubiquitous sources of parasitic signal, which have necessitated the development of specific approaches to realise background-free SRS.[25]

The main difference between CARS and CSRS processes is that they excite vibrational coherences that are complex conjugates of each other: first two pulses of the CARS process excite $|vn\rangle\langle v0|$ vibrational coherences in the ground electronic state, while those for the CSRS process generate $|v0\rangle\langle vn|$ coherences. As a consequence, the CSRS signal is detected on the low-energy side of the probe beam, where Raman features are often obscured by the presence of strong fluorescence background. This is, in essence, the main reason for the preference of CARS over CSRS in scientific community. In principle, both CSRS and CARS processes should lead to mostly equivalent information on Raman active vibrations, especially when the energies of the excitation beams are tuned far below any electronic transitions. However, in the case of resonantly enhanced RSS, where vibrational excitations on the excited electronic state are possible, the information contained in the CSRS signal can be different and complementary to that of CARS.[26] Moreover, a simultaneous detection of CSRS and CARS can also be beneficial for the enhanced detection sensitivity.[27]

Resonant excitation significantly enhances the sensitivity of a four-wave mixing experiment (including CARS and CSRS), but at the cost of added complications, such as generation of fluorescence, described above, and contributions from excited-state pathways. These can include vibrational modes on the excited electronic states, vibronic modes (where the electronic and vibrational degrees of freedom are strongly coupled), or electronic coherences, all of which can introduce additional peaks to the Raman spectrum. The challenge of identifying and separating coherences on ground or excited electronic states, as well as distinguishing between coherences of different nature (vibrational, vibronic or electronic) has been discussed extensively in relation to CARS[28] and MDES[29–31] experiments.

In the case of CARS and CSRS, separating excited electronic state vibrations from the ground electronic state vibrations has been achieved through control over several degrees of freedom: timings between three excitation pulses, polarizations, wavelengths, and detection windows.[28] Using MDES measurements the nature of coherences can be identified by comparing peak amplitudes and locations in a 3D spectrum which can be very difficult and time consuming to acquire. In the recent work, pathway selective MDES has been developed to access the specific parts of the 3D spectrum that can reveal the nature of coherences, without having to acquire the full 3D spectrum. The approach we use here is based on these pathway selective MDES experiments.[30,32–34]

With these measurements we demonstrate the detection of low-energy vibrational modes with resonant enhancement without the fluorescent background and NRB, and the ability to easily switch between different signal generation pathways. Specifically, we show a direct comparison between CSRS and CARS schemes, which allow us to identify excited-state vibrations, and isolate ground state contributions.

**Experiment**

Our experimental approach utilizes several ideas, developed within CARS and MDES. Specifically, we take advantage of an arrangement of the excitation beams in a box geometry, acquire spectral information in the time domain (i.e. Fourier transform spectroscopy), use heterodyne detection with phase cycling, and a pulse shaper to control the spectral and temporal properties of each individual pulse so that we can controllably access specific quantum pathways. Technical details on the experimental realization are described in the supplementary material, while the conceptual ideas are discussed below.

The box geometry (Fig. 1a,c), frequently used in CARS spectroscopy,[35,36] spatially separates the signal from the excitation pulses, providing reduced spectral background and greater selectivity of the signal generation pathways. In this geometry, the pump, the Stokes, and the probe beams with wavevectors $k_{pump}$, $k_{Stokes}$ and $k_{probe}$, respectively, are placed at three corners of a square. The signal of interest is emitted on the fourth corner of the square in the direction, $k_{CSRS} = - k_{pump} + k_{Stokes} + k_{probe}$ for the CSRS signal (or $k_{CARS} = k_{pump} - k_{Stokes} + k_{probe}$ for CARS), defined by the phase-matching condition. Spatial separation of the signal in this way allows a greater dynamic range and sensitivity of the detection.

In both CARS and CSRS spectroscopic schemes, the pump and the Stokes beams combine to excite the vibrational coherences (Fig. 1d,e) in either the ground or excited electronic state. The probe beam interacts with the vibrational coherences and is scattered into the signal direction with increased (CARS) or decreased (CSRS) energy. In time-domain measurements the oscillating phases of the vibrational coherences are mapped onto the signal phase which is measured as a function of the delay of the probe beam, $t_2$ (Fig. 1b,d). A Fourier transform of the data with respect to this time delay yields the spectral information on the vibrational modes. Such Fourier transform CARS spectroscopy has been applied previously[37,38] and has the advantage of being able to detect low-energy vibrations and having resolution limited only by the range of the delay. This relieves the need to have spectrally narrow pulses and/or filters with sharp edges. Measuring in the time domain can also help to remove NRB contributions,[39–41] which persist only while all three pulses overlap in time.

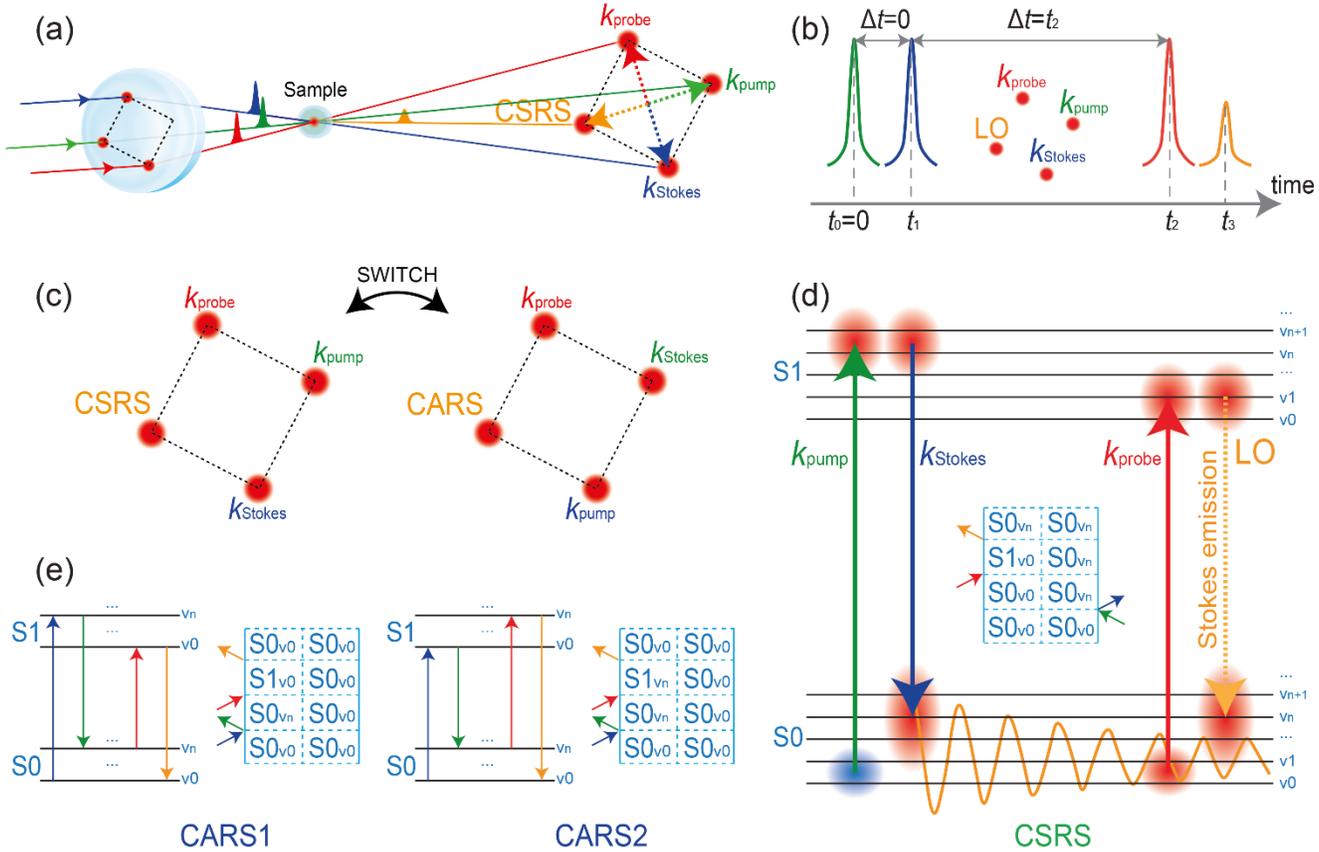

FIG. 1. Schematic representation of our pathway-selective CRS experiment. (a) Three spectrally shaped ultrashort laser pulses, arranged in the box geometry, are focused onto the sample to selectively excite different Liouville-space quantum pathways. (b) The first two pulses, the pump and the Stokes, arrive simultaneously, while the third, probe, pulse, delayed by the time $t_2$, generates a four-wave mixing signal, which interferes with the attenuated fourth pulse, referred to as local oscillator (LO). (c) Switching between different signal generation quantum pathways allows for the detection of CSRS and CARS signals in the same direction, defined by the wavevectors $k_{CSRS} = -k_{pump} + k_{Stokes} + k_{probe}$ and $k_{CARS} = k_{pump} - k_{Stokes} + k_{probe}$, respectively. In (d) and (e) Liouville-space pathway diagrams are illustrated alongside with the schematic representations of CSRS and CARS dipole transitions between electronic singlet states S0 and S1, having vibrational structure v0, v1, … . Two distinct CARS pathways, CARS1 and CARS2, are measured in the present work to distinguish between the excited-state and the ground-state vibrations.

To measure the phase of the signal electric field as well as the amplitude, we use heterodyne detection, which involves interfering the signal with a local oscillator (LO).[42] To record the signal we use a spectrometer and a CCD, where interference between the signal and the LO generates a spectral interferogram, from which we obtain the phase and the amplitude of the signal. By directly measuring the phase (as opposed to relying on interference between different signal contributions), we are able to get a unique solution to the Fourier transform, eliminating peaks that arise from linear combinations of the intrinsic vibrational modes. Another advantage of heterodyne detection is that the sensitivity of the experiment is further enhanced because the measured response depends linearly on electric field rather than intensity, as is normally the case.

One of the challenges of interferometric detection is that it requires interferometric phase stability between all pulses. This is achieved by generating the four beams using a two-dimensional spatial light modulator (2D-SLM) with a 2D phase grating; ensuring all beams are incident on common optics; or, where delays are introduced, they are split into pairs in a particular way that eliminates any phase jitter in the interferogram, (details can be found in [[34,43]] and in the supplementary material). These approaches have been developed particularly for multidimensional coherent electronic spectroscopy,[14,34] and have been applied for CARS spectroscopy as well.[20]

The challenges of heterodyne detection are offset by many advantages. In addition to those mentioned above, measuring the phase also allows phase cycling,[44–46] thus enabling significant suppression of contributions from undesired signals and enhancement of the signal of interest. In the simplest case phase cycling involves shifting the phase of one pulse by π, which also shifts the phase of the signal (and the resultant interferogram) by π. Subtracting this from the signal without any phase shift, doubles the signal, but removes any contributions (e.g. fluorescence) that do not have the same phase shift. In this case however, there remains a contribution from the interference of the beam that is phase cycled and the LO. To remove this contribution, and all others, we use an eight-step phase cycling, which involves 8 different combinations of pulses with phase shifts such that all that remains when combined in the right way is the interferogram of the signal and the LO. In this way, it is possible to eliminate all contributions from the pathways that do not depend on all four light-matter interactions, and enhance the desired signal (see the supplementary material for details). This eliminates, for example, fluorescence background, scatter from the excitation beams, and third-order signals from any other quantum pathways.

In the experiments reported here, the phase cycling is achieved by a 2D-SLM-based pulse shaper. The pulse shaper is configured so that each of the four beams are spectrally dispersed by a diffraction grating and a cylindrical lens in the horizontal direction and are incident on different vertical positions on the 2D-SLM matrix.[32] This configuration allows control of the spectral amplitude and phase of the pulses. In addition to phase cycling, the phase control allows optimal compression of the pulses to their Fourier-transform limit, ensuring a flat spectral phase over the whole bandwidth which helps with analyzing the spectral interferograms.

Control of the spectral amplitude allows the pulses to be shaped so that specific pathways can be excited. In particular, we begin with broad pulses (covering spectral region of ~760-870 nm and having FWHM of ~70 nm) and shape the spectra of the individual beams, as shown in Fig. 2, after which we compress all pulses down to ~65 fs. Making the pump and the Stokes pulses non-degenerate ensures any detected signal arises from coherence pathways. In the context of non-resonant Raman spectroscopy, this may seem trivial since all vibrational pathways involve a coherence between vibrational levels. However, in resonant experiments where electronic excitations also occur, a part of the signal can arise from pathways involving excited (or ground) state populations. Indeed, this contribution can dominate the signal and swamp the desired vibrational pathways. Shaping the probe pulse further allows us to select specific resonant pathways and, as a consequence, to generate a signal that is spectrally well separated from the excitation pulses. Pulse shaping provides a flexible and straightforward way of changing between different pathways excited. Instead of tuning the wavelength of different optical parametric amplifiers (OPAs) or changing filters, all that is required is selecting between different amplitude modulation functions applied to the pulse shaping 2D-SLM. In particular, this allows us to easily switch between CARS and CSRS schemes (Fig. 1c and Fig. 2), both of which can affect ground and excited state vibrational coherences differently. This approach of incorporating pulse shaping in $3^{rd}$ order spectroscopy has been previously adopted in MDES in order to isolate coherence pathways[14,33,34] and is very similar to the approach of the present work.

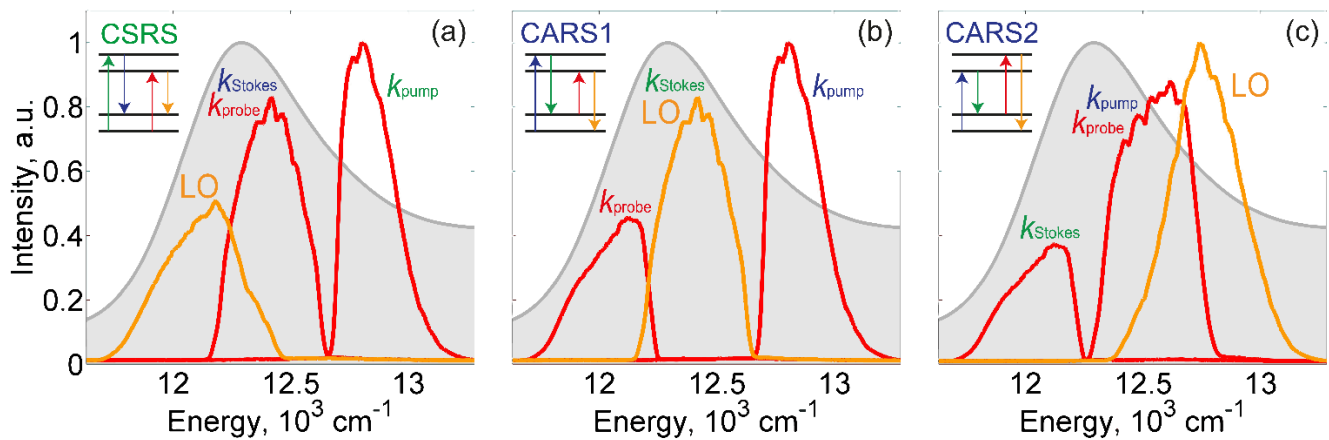

FIG. 2. The shaped pulse spectra of the pump, the Stokes, the probe (red) and the LO (orange) beams used for the (a) CSRS, (b) CARS1 and (c) CARS2 signal generation pathways. The absorption spectrum of IR-813 is shown shaded in grey on the background. For convenience the corresponding energy level diagrams and the dipole transitions are shown in insets.

To demonstrate this technique and its capabilities we measure the spectra of near-infrared (NIR) cyanine dyes as test samples (see the supplementary materials for more description of the samples). Here we present the results for the NIR cyanine dye IR-813; similar results for other dyes, IR-806 and IR-140, are presented in the supplementary material. The absorption and fluorescence spectra of IR-813 are shown in Fig. 3 together with the broadband laser spectrum, which is tuned to overlap most of the absorption band. In order to selectively excite specific coherent Raman pathways, as described above, the pump, the Stokes, the probe and the LO pulses are spectrally shaped as illustrated in Fig. 2. Three different configurations of spectral profiles of the pulses are considered: one configuration for driving CSRS quantum pathways and two different configurations for driving CARS (CARS1 and CARS2) quantum pathways. In each case, the pump and the Stokes spectra are shaped so that there is negligible spectral overlap between the two, ensuring that the signal predominantly arises from the coherence pathways (see the supplementary material for more details on the effect of having greater spectral overlap between the first two pulses). The range of vibrational energies that can be excited is limited by the bandwidth of and energy separation between the pump and the Stokes pulses; namely, it is mostly defined by the spectral cross-correlation of the pump and the Stokes pulses, which is imprinted into the relative intensities of the detected Raman features. The probe and the LO spectra are chosen so that the signal overlaps with the spectrum of the LO, which leads to the additional scaling of the relative signal intensities by the spectral shapes of the last two pulses. To reconstruct the relative strengths of the detected Raman peaks it is then necessary to take into account the spectral shapes of all four pulses (see more details on the intensity calibration in the supplementary material).

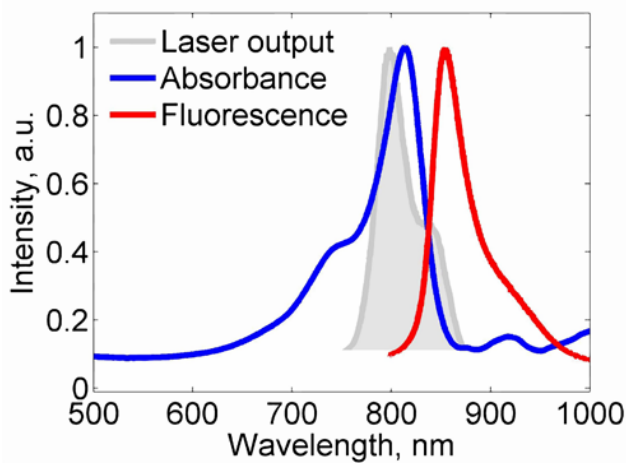

FIG. 3. Normalized absorbance (blue) and fluorescence emission (red) spectra of IR-813, and the broadband laser spectrum (grey) on the background.

The quantum pathways being excited in these measurements are shown in Fig. 4. In both the CSRS (Fig. 4a,b) and CARS1 (Fig. 4c,d) experiments the combination of pulse energies, set by the pulse shaper, allows resonant excitation of vibrational coherences in both ground S0 (Fig. 4a,c) and excited S1 electronic singlet states (Fig. 4b,d). However, in the CARS2 experiment with similar resonant enhancement, the pathway involving vibrational coherences in the excited electronic state S1, is expected to have reduced amplitude. This is because to access the same transitions as in the CSRS and CARS1 schemes, the initial state has to be the $|S0_{vn}\rangle$ state, as shown in Fig. 4f. This state will have some thermal population at equilibrium, but for modes >200 cm$^{-1}$, this population will be minimal. Therefore, if there is a substantial contribution to the signal from vibrational modes on the excited electronic state, S1, a significant difference in the detected spectra would be expected between CARS2 and the other two experiments. Other pathways beginning from the ground $|S0_{v0}\rangle$ state are possible in the CARS2 scheme, but these excited-state absorption pathways are equally possible in the CARS1 and CSRS schemes.

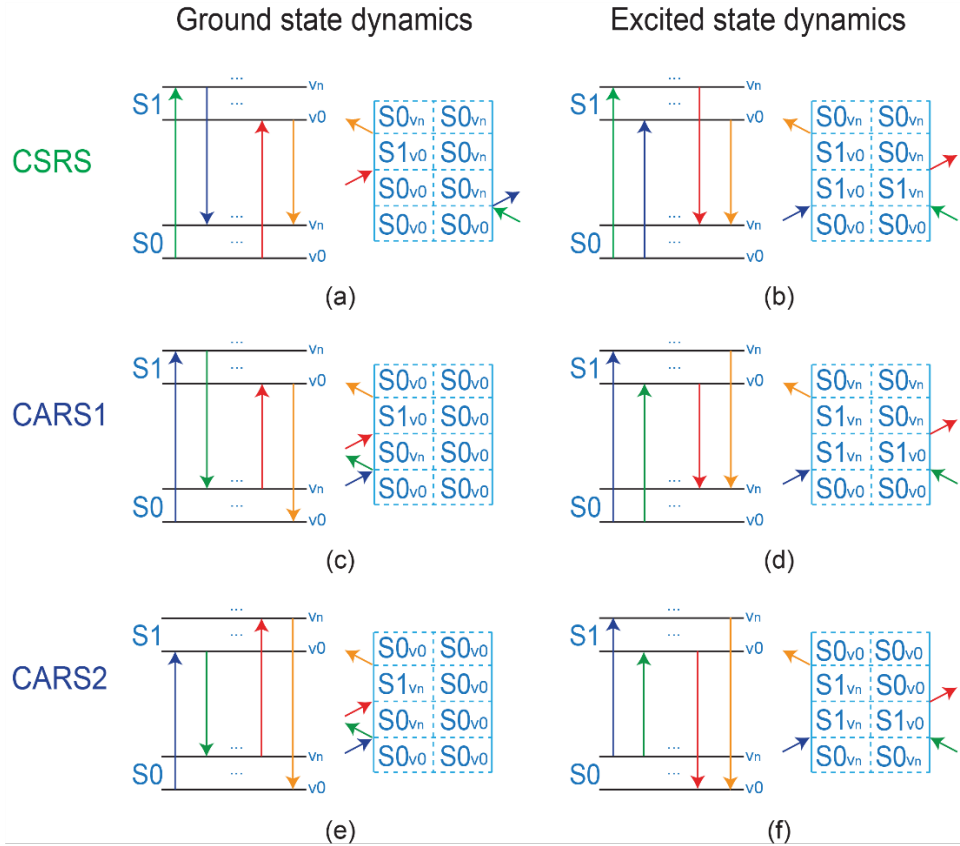

FIG. 4. The generic pathways showing the energy levels and Feynman-Liouville diagrams for the different pulse sequences used. The pathways exciting vibrational coherences in the ground electronic S0 state are shown in (a), (c), and (e), and the pathways leading to vibrational coherences in the excited electronic state S1 are illustrated in (b), (d), and (f).

**Results and Discussion**

The Raman intensity spectra, acquired from our measurements, are represented in Fig. 5, where the relative intensities of spectral features have been calibrated and the effects of the different excitation pulse spectra have been taken into account. The positions of Raman peaks, above 200 cm$^{-1}$ and their widths are in excellent agreement with surface-enhanced Raman scattering (SERS) spectra, reported previously.[47,48] The present measurements, however, extend the lower detection limit below 200 cm$^{-1}$ (down to ~30 cm$^{-1}$) and reveal additional peaks (all detected Raman signatures are listed in the Table 2 in the supplementary material). The Raman line shapes obtained are slightly asymmetric due to the initial thermal population of the vibrational modes below ~200 cm$^{-1}$ (thermal energy) and asymmetries in the confinement potential.

The comparison of the Raman signals, generated by driving electrons through CSRS, CARS1 and CARS2 quantum pathways, is shown in Fig. 5a, where all three spectra were acquired sequentially in nearly identical experimental conditions.

There is a number of differences between the CSRS and CARS spectra observed. First, the CARS features exhibit lower intensities than those of CSRS: the intensity of the CARS1 signal is lower than the intensity of the CSRS peaks by a factor of ~2. More strikingly, the features of the CARS2 spectrum are ~1200 times weaker than those of the CSRS spectrum. These numbers take into account differences in integration times and in spectral shapes of the excitation pulses. Since CSRS and CARS1 schemes are driven with the same laser pulses, which interact with the exact same transitions, it is expected that the strengths of the two signals should be of the same order of magnitude. The CARS2 pathway involving the vibrational modes on S0 (Fig. 4e) also involves the exact same transitions as CARS1 and CSRS, but as discussed above, the pathway involving vibrational coherences on S1 (Fig 4f) are expected to be significantly less efficient than the equivalent pathways in CARS1 and CSRS (Fig. 4b). From this stark difference we conclude that the dominant contribution to the differences in the signals measured here is vibrational coherences on the S1 excited electronic state. This is further confirmed by the observation that the difference between CARS1 and CARS2 signal amplitudes appears to decrease as the energy of the mode decreases, and the thermal population of these modes at equilibrium increases (see e.g. the peak at $138.5 \pm 1.3$ cm$^{-1}$).

Another observed difference between the CSRS and CARS spectra is that the spectral tails of individual Raman features are on different sides: in the CARS spectra the tails are observed on the low-energy side of the spectral features, whereas in the CSRS spectrum the tails are observed on the high-energy side (see also the amplitude spectra in the supplementary material). This is because the Liouville subspace describing CARS quantum pathways has been folded into its conjugate half to represent both CSRS and CARS spectra in the same spectral domain (with positive Raman shifts).

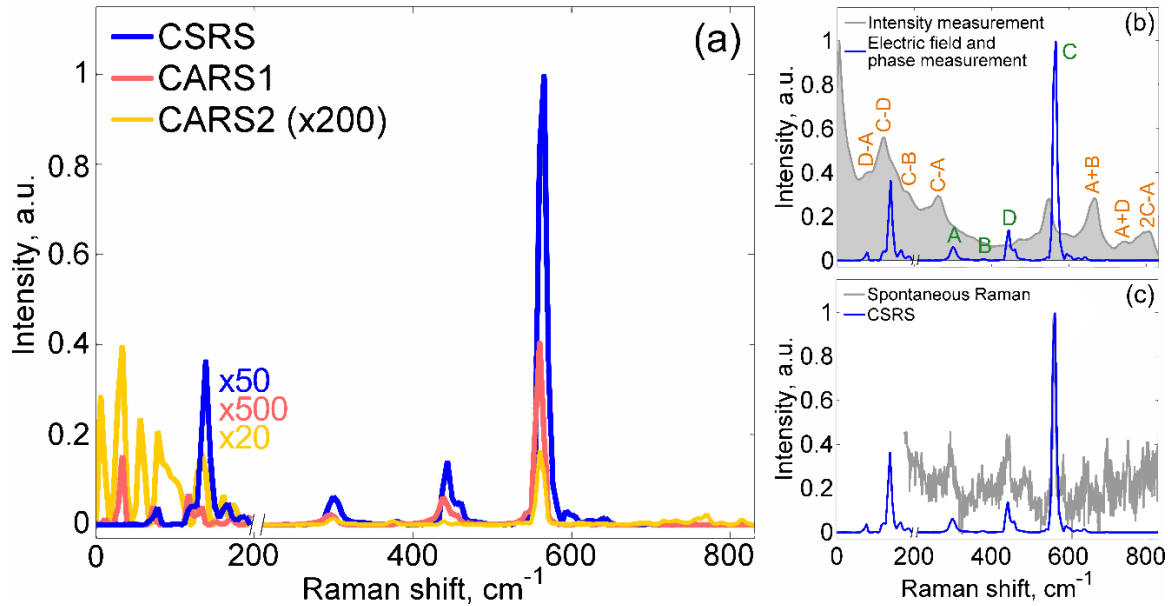

FIG. 5. (a) Comparison of the 1D CSRS, CARS1 and CARS2 (scaled by a factor of 200 in the >200 cm$^{-1}$ region) calibrated intensity spectra. The most intense CARS2 feature is ~1200 times weaker than the corresponding CSRS peak. The Raman features below 200 cm-1 are scaled by a factor of 50 (CSRS), 500 (CARS1) and 200x20 (CARS2). (b) Comparison of the calibrated 1D CSRS intensity spectra (blue) of the IR-813 with analogous spectra, generated by time-resolved intensity measurements (grey): intensity measurements lead to congested spectra, having additional peaks, corresponding to linear combinations of intrinsic Raman modes. (c) Comparison of the calibrated 1D CSRS spectra (blue) with the spontaneous Raman lineshapes (grey), obtained by the conventional Raman spectroscopy apparatus (Renishaw confocal Raman microscope) using 488 nm excitation wavelength. The fluorescence background in the spontaneous measurements was subtracted. The integration time was set to 50 sec to observe noticeable signal. See more information on these measurements in the supplementary material.

To highlight the importance of the interferometric detection of the phase information, we compare our data to the equivalent case where only intensity is measured (Fig. 5b) by taking the modulus square of our data in the time domain $t_2$. This keeps the benefits of phase-cycling and interferometric detection for removing the various background signals, however, introduces additional spurious frequency components. In the case of intensity measurements (shaded in grey in Fig. 5b), when the phase information is absent, the acquired Raman spectra are congested with spectral features that are linear combinations of the peaks A, B, C, and D. This is especially problematic in the low-energy region where rectified Raman features (differences of fundamental Raman modes, including non-oscillating contributions) dominate over intrinsic vibrational modes, making interpretation of the Raman modes challenging.

Fig. 5c shows a further comparison of the CSRS spectrum with the results acquired by means of commercial spontaneous Raman spectroscopy instrument (see the supplementary material for details). As expected for a resonance-enhanced coherent Raman spectrum, the signal-to-noise ratio is significantly higher, by a factor of ~250. This is based on the RMS noise in the data and does not take into account the different acquisition parameters, such as the 50 seconds integration time required for the spontaneous Raman measurements and different volumes probed. In our CRS measurements the integration time for each step was 30 ms (CSRS), 20 ms (CARS1), and 8 ms (CARS2) with overall integration times per spectrum being ~39 sec, ~26 sec and ~10 sec, respectively. The comparison also shows that spontaneous Raman features are narrower than those acquired in the CSRS/CARS measurements: the FWHM of the intense Raman feature obtained from the spontaneous Raman experiment was measured to be ~6.5 ± 0.6 cm$^{-1}$ (compared to ~12.5 ± 2.6 cm$^{-1}$ for the case of CSRS, ~10.9 ± 2.6 cm$^{-1}$ for the case of CARS1, and ~11.2 ± 2.6 cm$^{-1}$ for the case of CARS2). This may be because in the CSRS/CARS measurements we are probing the vibrations on the excited electronic state, which are typically broader, whereas the spontaneous Raman probes the vibrations on the ground electronic state. However, other experimental effects such as power broadening or the longer excitation wavelength may be artificially broadening the spectrum.

**Conclusion**

In conclusion, the approach described here combines several techniques that have been used for coherent Raman measurements previously in a manner that draws on recent developments in multidimensional coherent electronic spectroscopy. This approach allows detection of resonance-enhanced CSRS/CARS spectra with Raman shifts down to ~30 cm$^{-1}$, high sensitivity, and without the presence of fluorescence and non-resonant backgrounds. The intensity of the Raman features far below 200 cm$^{-1}$ could be enhanced by precise crafting of the spectral shapes of the excitation pulses. The ability to control precisely the spectral shape of each of the pump, Stokes, probe and LO pulses gives added flexibility for coherent Raman spectroscopy, and allows straightforward interchange between CSRS and CARS experiments, and between different quantum pathways in general. This is particularly useful in the cases, where the two schemes provide complementary information, such as in resonant experiments where excited state vibrations can be probed.[49] Indeed, recent measurements have shown the ability to separate contributions from vibrational modes on ground and excited electronic states in resonant Raman experiments by shifting broadband pulses above and below resonance.[50–52] Our ability to impart even greater control on the pulse spectra and available pathways provides even greater capacity to separate these contributions. Finally, the flexibility of the experiment leaves it well placed to extend these measurements to higher order experiments, such as the 2D resonant Raman[53] and 2D-Raman 2D electronic[54,55] spectroscopies recently reported.

**Supplementary material**

See the supplementary material for more details on the experimental apparatus, the dependence of the detection window on spectral overlap of the excitation pulses, the details on the intensity calibration procedure for the correction of relative intensities of the observed vibrational modes, CSRS measurements of IR-806 and IR-140, and spontaneous Raman measurements.

**Acknowledgments**

This work was supported by the Australian Research Council (FT120100587).

# Supplementary materials:
# Background-free time-resolved coherent Raman spectroscopy (CSRS and CARS): heterodyne detection of low-energy vibrations and identification of excited-state contributions

Pavel Kolesnichenko,[1,2] Jonathan O. Tollerud,[1] and Jeffrey A. Davis[1,2,a]

[1]*Centre for Quantum and Optical Science, Swinburne University of Technology, Melbourne, Victoria 3122, Australia*

[2]*Centre for Future Low-Energy Electronics Technologies, Swinburne University of Technology, Melbourne, Victoria 3122, Australia*

**Contents**

1. Technical details on experimental realization
2. 8-step phase cycling
3. Samples
4. Intensity calibration of CSRS, CARS1 and CARS2 Raman spectra
5. Dependence of the detected Raman spectra on the spectral overlap between the pump and the Stokes excitation pulses
6. Raman spectra of IR-813, IR-806 and IR-140
7. Spontaneous Raman measurements
8. Anti-Stokes-to-Stokes ratio in the limit of low energy vibrations

## 1. Technical details on experimental realization

A simplified representation of our apparatus, which is based on previous approaches in MDES,[1–3] is depicted in Fig. S1. The mode-locked Ti:Sapphire oscillator (KMLabs Inc.), pumped by 532 nm continuous wave (cw) laser beam, provides the pulse train at 92 MHz repetition rate. The spectral bandwidth of the pulses was ~70 nm, centered at ~800 nm. The beam, indicated by Ⓐ in Fig. S1, was focused onto a beam shaper (incorporating a 2D (512x512 pixels) programmable liquid crystal on silicon spatial light modulator (SLM), from Boulder Nonlinear Systems, Inc. (BNS)), which was programmed to generate a square diffraction pattern Ⓑ. The beam mask allows only the radiation, diffracted into the first diffraction order, to pass through, thus leaving only four beams, arranged in the box geometry rotated by 23 degrees Ⓒ. The advantage of using an SLM rather than a simple 2D diffraction grating (DG) lies in its ability to arbitrarily shape the beam, allowing to choose amongst several phase matching beam geometries as well as to add additional beams to investigate higher order coherent processes on demand. The beam shaper also allows tuning intensity of each outgoing beam, assisting, particularly, in attenuation of the local oscillator (LO). After the beam shaper the four beams propagate towards a 4*f* pulse shaper, consisting of a DG (1200 grooves/mm), a cylindrical lens (CL) and an additional SLM (BNS). With the four beams separated vertically and spectrally dispersed horizontally on the SLM, Ⓔ, the phase of each frequency component of each beam can be controlled independently. In addition to the spectral phase modulation, a vertical saw-tooth phase pattern was applied to the pixels of the SLM2, efficiently simulating a diffraction grating. This modulation shifts the beam downwards so that they can be directed to the delay stage (DS) area by a pick-off mirror POM. By changing the amplitude of the saw-tooth pattern the spectral amplitude of the pulses can also be controlled.

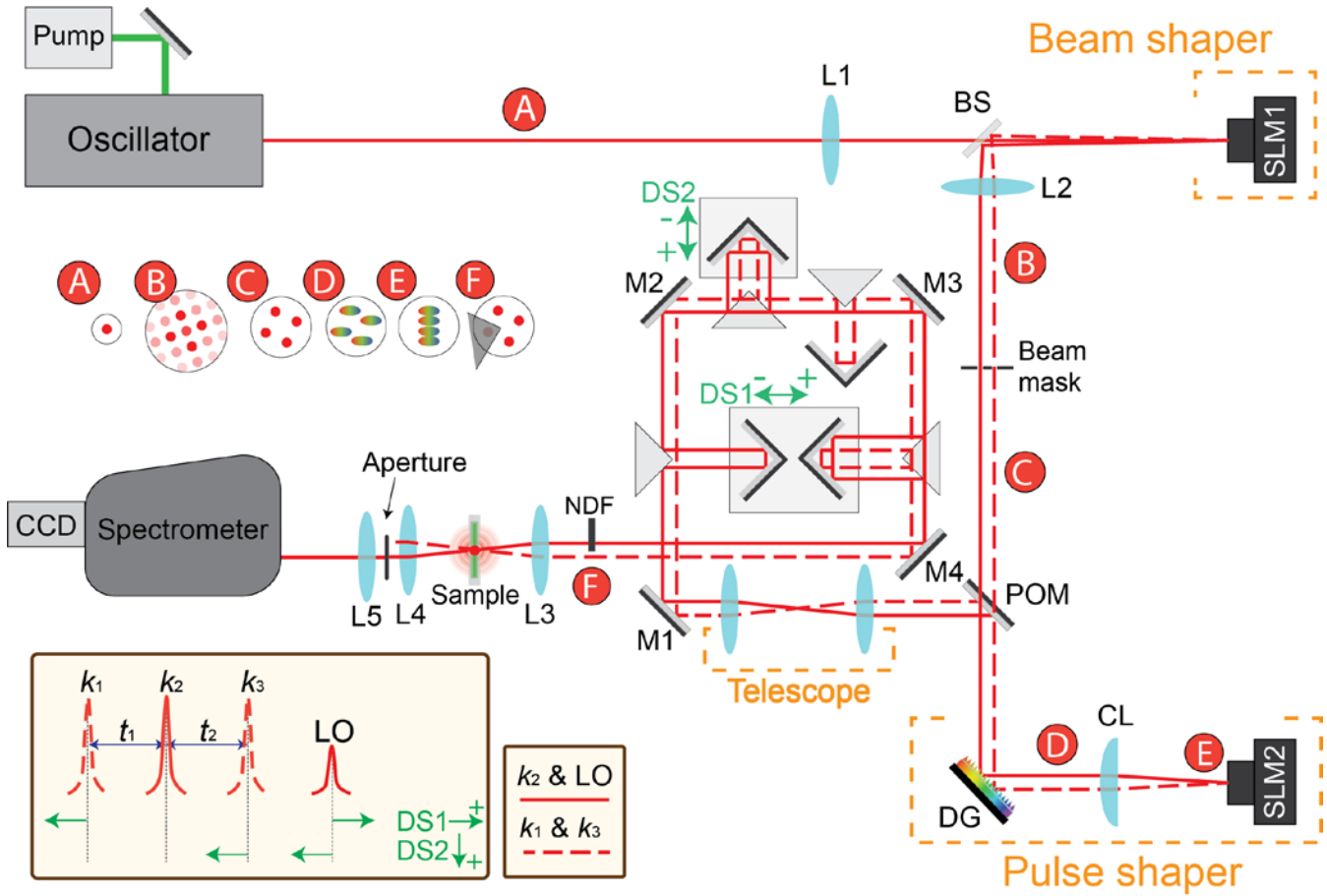

FIG. S1. Schematic representation of the CSRS/CARS apparatus.

The pump pulse was shaped to have higher energies than both Stokes and probe pulses, with the spectral peak shifted by ~400 cm$^{-1}$ to cover few ground state vibrational levels. The wavelength range covered by each shaped excitation pulse was ~27 nm, with full width at half maximum (FWHM) $\Delta\lambda \approx 20 \pm 5$ nm. To reduce the contribution to the signal from population pathways, the spectral overlap between the pump and the Stokes pulses was minimized. The effects of amplitude shaping on the resulting spectra are discussed in more details in section S5. All pulses were compressed to their Fourier transform limit down to Δτ ≈ 65 fs, making them optimal for the four-wave-mixing (FWM) processes. The phase modulations required for pulse compression were also applied by the pulse shaper.

After the pulse shaper, four beams enter the DS area, with each delay line (DL) consisting of a prismatic mirror and a retroreflector. The delay $t_1$ between the pump and the Stokes pulses can be introduced by the DS1, while the delay $t_2$ between the pump and the probe pulses can be controlled by the DS2. For the CARS and CSRS experiments we kept the pump and the Stokes pulses overlapped in time, while delaying the probe pulse, which allowed tracking the dynamics of zero-quantum (0Q) coherences (i.e. the Raman coherences) in the samples under consideration. The beams were split in pairs so that phase fluctuations normally associated with travelling different paths, were cancelled in the detected interferogram.[4,5] This can be understood as follows. The phase of the signal $\varphi_{sig.}$ will depend on the phases $\varphi_{pump}$, $\varphi_{Stokes}$, $\varphi_{probe}$, and $\varphi_{LO}$ of all pulses in accordance with the relation

$$\phi_{sig} = -\phi_{pump} + \phi_{Stokes} + \phi_{probe} - \phi_{LO}. \tag{1}$$

If a phase fluctuation Δφ were to occur in those terms of the relation (1) that have opposite signs, then the fluctuation would cancel out. Only four such combinations of beams are possible, leading to the four DLs, as shown on the Fig. S1. This minimizes the phase fluctuations, taking place due to the jitter of the prisms and the retroreflectors. The phase stability of the entire setup was measured to be better than λ/300.[2]

After the DS area, all beams propagate through a 7.5 cm focal length lens with the numerical aperture NA = 0.6 at 800 nm wavelength, focusing all beams into a spot of a diameter $d ≈ 60$ μm. The average power in each beam before the lens was $P ≈ 1$ mW. The LO was attenuated by the beam shaper by a factor of 2.2 and then further by the neutral density filters (NDF) with overall optical density OD = 4.8. Samples were placed in the fused silica sample holder with the spacing width $l = 100$ μm. The signal was frequency-resolved by a spectrometer (IsoPlane® SCT 320) coupled to a CCD camera (PIXIS, Princeton Instruments), giving rise to the *detection energy* axis in the acquired 2D spectral maps

To acquire simultaneously information on the electric field amplitude and the phase of the signal, we took an advantage of the interferometric (heterodyne) detection. Derived from the same laser source, the LO was set as a reference pulse to interfere with the emitted signal, producing interference fringes. The LO was delayed by ~1.2 ps, allowing us to unambiguously discern fringes in the detected spectral interferogram and eliminate scatter contributions. To reveal the weak signal imprinted into the significantly larger LO and reduce all other spectral perturbations (e.g. scatter, and fluorescence background), while simultaneously increasing the signal-to-noise (S/N) ratio, we applied 8-step phase cycling, implemented through the pulse shaper. The interferograms were detected for each delay step $t_2$ (with the increment of 20 fs) from -0.2 ps to 3 ps. The signal's amplitude was acquired as a function of time delay $t_2$ and the CCD's detection energy. For a particular experiment, the integration time was set to keep the detected LO intensity close to the limit of the dynamic range of the CCD. The overall acquisition time, thus, varied for different measurements, and can be approximately estimated by multiplying the integration time, the number of phase cycling steps (eight), the number of averages (five) and the number of time steps (161). The signal processing was done by first Fourier transforming the signal with respect to the *detection energy ($\omega_3$)*. The data is then windowed to select the physical time ordering for the signal and filter out the residual scatter, the fluorescence background and much of the NRB. The delay between the signal and LO is removed and the data is Fourier transformed with respect to both $t_2$ and $t_3$ to generate a 2D spectrum. The Raman spectra shown in Fig. 5 are then obtained by integrating across the $\omega_3$ axis.

## 2. 8-step phase cycling

A CCD camera is an intensity detector and, thus, detects the following signal:

$$S \sim \left| E_{LO} e^{i\phi_{LO}(\omega)} + E_{sig} e^{i\phi_{sig}(\omega)} + E_{pump} e^{i\phi_{pump}(\omega)} + E_{Stokes} e^{i\phi_{Stokes}(\omega)} + E_{probe} e^{i\phi_{probe}(\omega)} \right|^2 =$$

$$I_1 + I_2 + I_3 + I_4 + I_5, \tag{2}$$

where

$$I_1 = E_{LO}^2 + E_{sig}^2 + \sum_n E_n^2, \tag{3}$$

$$I_2 = \sum_n \sum_{m}^{m \neq n} E_m E_n \cos(\phi_m - \phi_n), \tag{4}$$

$$I_3 = 2 \sum_n E_{sig} E_n \cos(\phi_{sig} - \phi_n), \tag{5}$$

$$I_4 = 2 \sum_n E_{LO} E_n \cos(\phi_{LO} - \phi_n), \tag{6}$$

$$I_5 = 2 E_{LO} E_{sig} \cos(\phi_{LO} - \phi_{sig}). \tag{7}$$

where $n, m = \{pump, Stokes, probe\}$.

For each time step $t_2$, we used 8-step phase cycling, implemented by the pulse shaping SLM. This allows to eliminate $I_{1-4}$ and amplify $I_5$, which carries the signal of interest. The eight steps of the procedure are summarized in the Table 1, where S1,...,S8 are the spectral interferograms as described in Eq. (2), and 0 or π are the phase shifts applied to the corresponding pulses.

TABLE 1. Eight steps of the phase cycling procedure.

| Step | pump | Stokes | Probe | LO |
|------|------|--------|-------|-----|
| S1 | 0 | 0 | 0 | 0 |
| S2 | π | 0 | 0 | 0 |
| S3 | 0 | π | 0 | 0 |
| S4 | π | π | 0 | 0 |
| S5 | 0 | 0 | 0 | π |
| S6 | π | 0 | 0 | π |
| S7 | 0 | π | 0 | π |
| S8 | π | π | 0 | π |

The signal is then determined as the following combination of the resulting interferograms, which cancels all but $I_5$, as required:

$$E_{sig} \sim [(S2-S1)-(S4-S3)] - [(S6-S5)-(S8-S7)] = -16 E_{LO} E_{sig} \cos(\phi_{LO} - \phi_{sig}). \qquad (8)$$

## 3. Samples

For our measurements, we used methanol solutions of NIR cyanine laser dyes IR-813, IR-806 and IR-140 (Sigma-Aldrich). These dyes are generally represented by a chain of odd number of methine groups (CH=) with two nitrogen atoms at the ends of the chain, which in turn are parts of heterocyclic aromatic molecules (Fig. S2). The conjugated methine groups provide $2p_z$ orbitals[6] that give rise to the first singlet transition S0 → S1. The excitation pulses together with the LO were tuned to be resonant with the first transition S0 → S1. The absorption and fluorescence spectra of each of the dyes are represented on the Fig. S3.

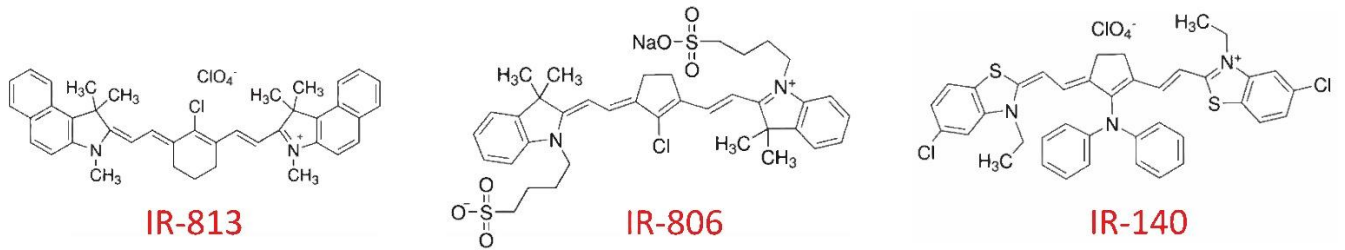

FIG. S2. Chemical representation of the cyanine laser dyes IR-813, IR-806 and IR-140 (Sigma-Aldrich).

The choice of these dyes as the test samples was dictated by several reasons. First of all, the absorption spectra of the dyes overlap the spectral output of the laser, facilitating the resonant enhancement of the CSRS signal. Second, the samples have active Raman modes in our detection region. Finally, they produce strong fluorescence emission in the detection range of our technique (~30–820 cm$^{-1}$), which makes conventional spontaneous Raman measurements virtually impossible (see Fig. S9). Moreover, the fluorescence lifetimes of the samples exceed few hundreds of ps,[7] which is much longer than vibrational coherence times of the dyes, making the presence of the fluorescence background in the time domain nearly constant during our time-resolved measurements, allowing to test our ability to minimize the fluorescence background in the detected spectra.

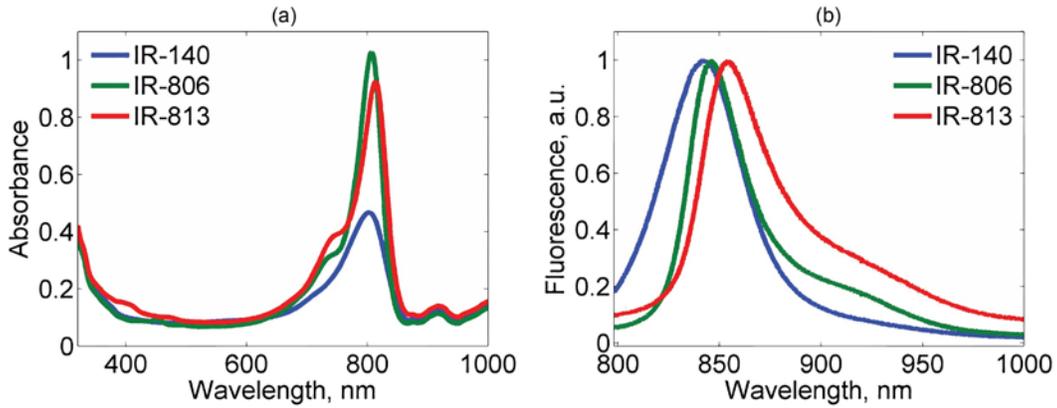

FIG. S3. (a) Absorbance and (b) normalized fluorescence spectra of three NIR cyanine dyes IR-813, IR-806 and IR-140.

The concentration of the dyes in the solutions was set to ~10 μM (OD ~0.01) and the volume, probed in the coherent Raman measurements, was ~0.2 nL.

4. Intensity calibration of CSRS, CARS1 and CARS2 spectra

We estimated the accessible detection window and spectral responsivity for each of the pathways using the spectral shapes of the pulses shown in the Fig. 2 of the main text. The overall shape can be defined by three parameters $\gamma_{LO}$, $\gamma_{pump*Stokes}$ and $\gamma_{probe}$ (see Fig. S4d), which are ultimately defined by the spectral widths of all participating pulses. The first two pulses excite the vibrational coherences with energies given by the energy difference between the pump and Stokes pulses. The distribution of energies accessible matches the cross-correlation of the pump and the Stokes spectra and defines the detection range $\gamma_{pump*Stokes}$ along the vertical axis (*Raman shift*). The horizontal extent $\gamma_{probe}$ of the individual Raman features along the *detection energy* axis is defined by the spectral bandwidth of the probe pulse. The measured response is a spectral interferogram between the signal and the LO, and thus is dependent on the product of the signal and LO amplitudes. The range $\gamma_{LO}$, which is the width of the LO, thus provides an additional windowing of the signal along the horizontal axis. The 2D response functions shown in Fig., S4 were obtained by taking the cross-correlation of the pump and Stokes spectra; taking the cross product of this with the spectrum of the probe; and multiplying by the LO spectrum along the detection energy axis. The overall responsivity across the Raman spectra were then determined by integrating across $\omega_3$, as shown in Fig. S5.

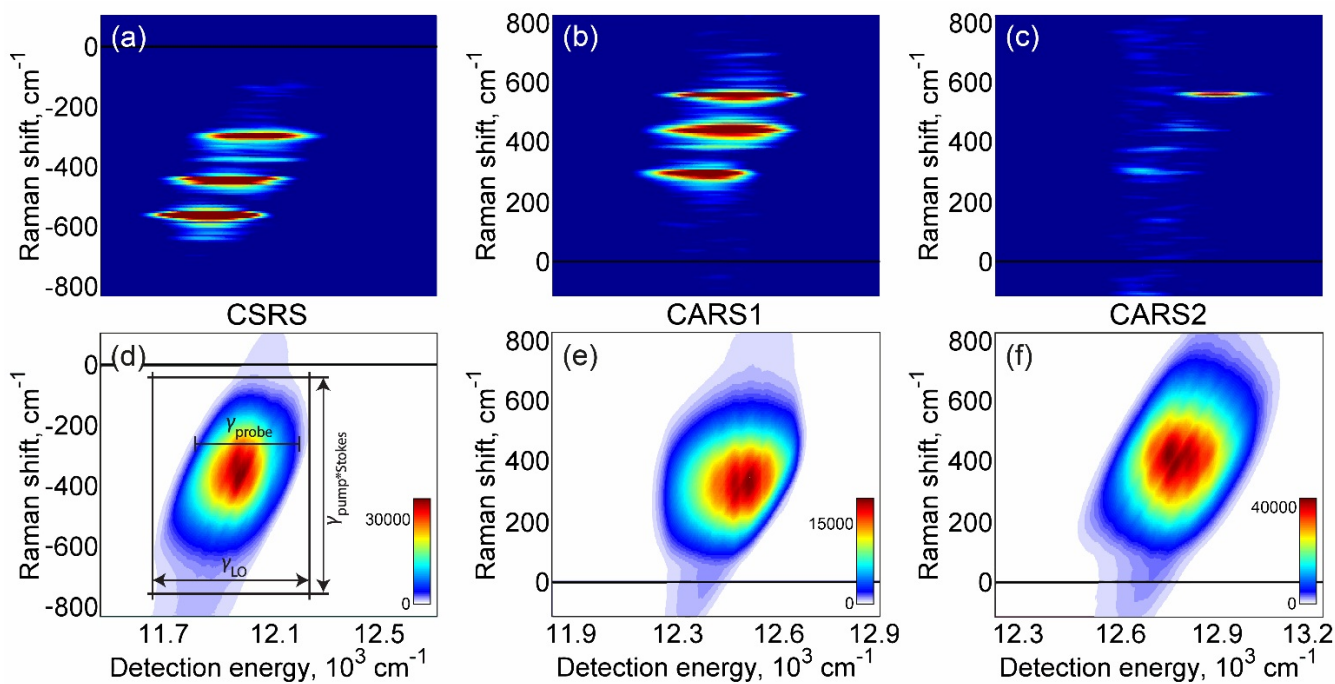

FIG. S4. (a-c) Acquired 2D spectral maps, correlating detection energy with observed Raman shifts, and (d-c) estimated 2D spectral response functiond for the cases of driving IR-813 dye molecules through (a,d) CSRS, (b,e) CARS1 and (c,f) CARS2 signal generation quantum pathways. The integration times were set to be 30 ms, 20 ms, and 8 ms for the CSRS, CARS1, and CARS2 measurements, respectively. The detection window parameters $\gamma_{pump*Stokes}$, $\gamma_{probe}$ and $\gamma_{LO}$ are delineated in (d). The black line marks the level of zero-energy Raman shift. The color scale limits in (a-c) are kept constant for the sake of direct comparison.

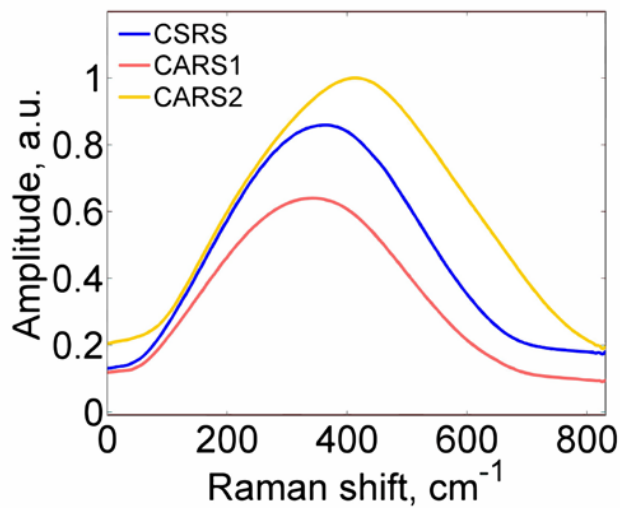

FIG. S5. Spectral response curves of the CSRS (blue), CARS1 (light red) and CARS2 (yellow) measurements, obtained by integrating corresponding 2D reponse along the horizontal (*detection energy*) axis.

The calibrated Raman amplitude spectra were obtained by dividing the measured spectra by the corresponding integrated responsivity curve (Fig. S6b). Finally, 1D raw (Fig. S6c) and calibrated (Fig. S6d) Raman intensity spectra are obtained by squaring corresponding 1D Raman amplitude spectra.

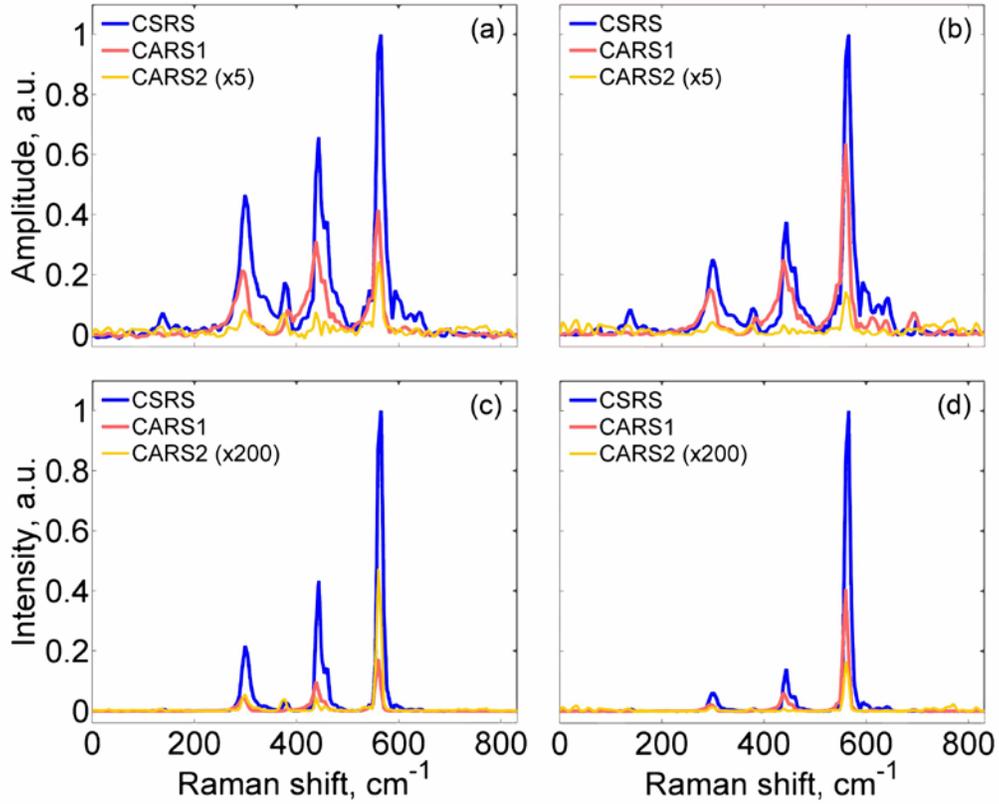

FIG. S6. (a) 1D Raman amplitude spectra of the CSRS (blue), CARS1 (light red) and CARS2 (yellow) measurements, obtained by integrating corresponding 2D spectral maps along the horizontal (*detection energy*) axis, and (b) calibrated 1D Raman amplitude spectra. (c,d) 1D Raman intensity spectra, obtained by squaring 1D Raman amplitude spectra in (a) and (b), respectively. In the case of CARS2 measurements, the amplitude spectra are scaled by the factor of 5, and corresponding intensity spectra are scaled by a factor of 200 for better comparison.

## 5. Dependence of the detected Raman spectra on the spectral overlap between the pump and the Stokes excitation pulses

We measured the dependence of the Raman spectra of the IR-813 on the spectral overlap between the pump and the Stokes pulses. The more the spectral overlap between the pulses, the more intense the contribution from the degenerate (population) quantum pathways becomes (Fig. S7), giving rise to a peak at $\omega_2 = 0$, corresponding to the component that doesn't oscillate. This is confirmed by the elevated amplitude of the detection windows in the low-wavenumber region (Fig. S8). It becomes evident that a strong zero-frequency peak introduces additional background, secondary side-lobes and noise.

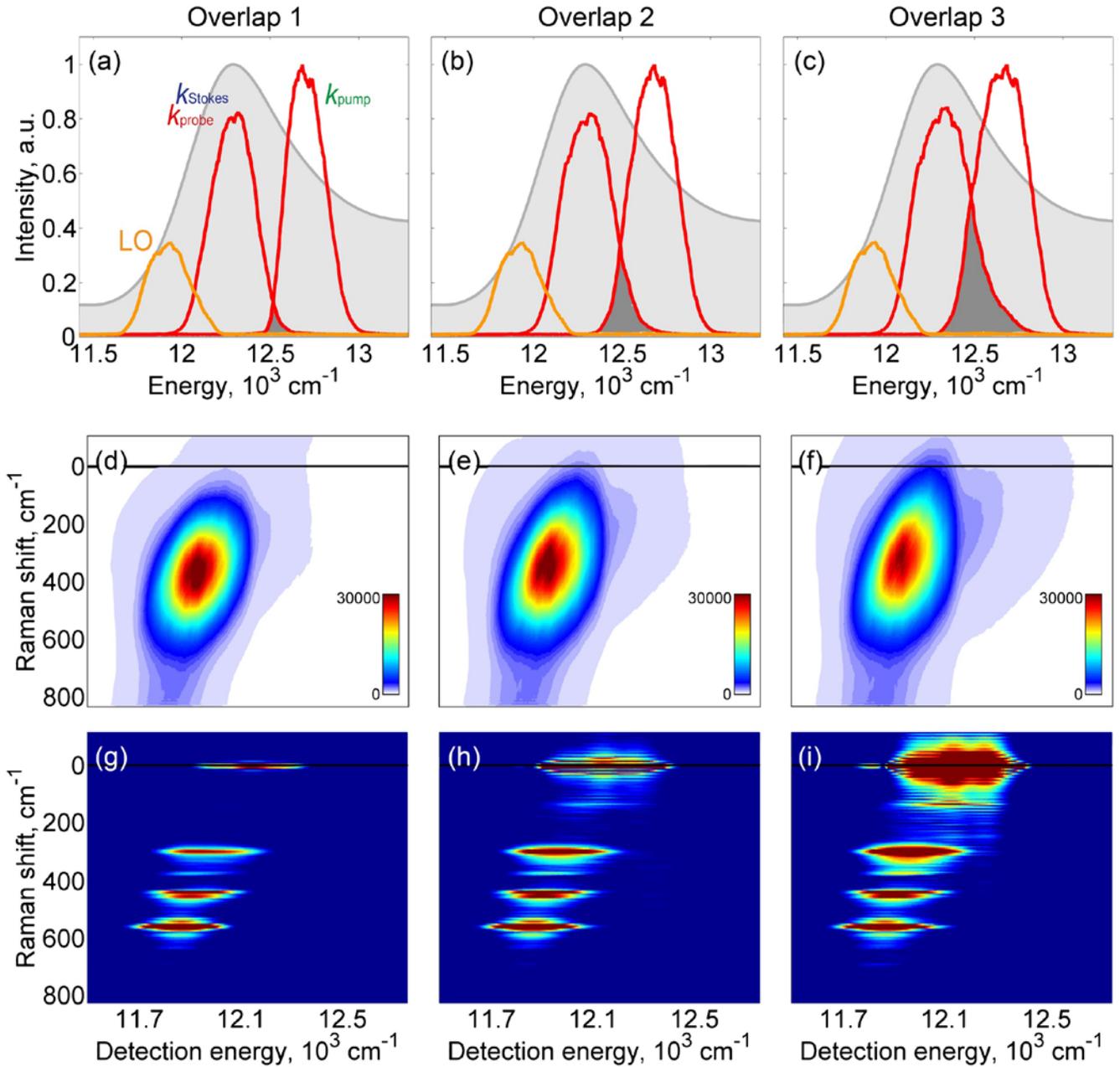

FIG. S7. Dependence of detection windows (d-f) and 2D spectral maps (g-i) on the amount of spectral overlap between the pump and the Stokes excitation pulses (a-c). In (a-c) absorption spectrum of the IR-813 is shown in light grey. Left column (a,d,g) corresponds to the case *overlap1* of a minor overlap between the pump and the Stokes spectral shapes; middle column (b,e,h) represents the case *overlap2* of an intermediate overlap between the pump and the Stokes pulses; right column (c,f,i) shows the case *overlap3* of a major overlap between the pump and the Stokes pulses. The limits of color scale in (g-i) are kept constant for the sake of direct comparison. The black line in (d-i) marks the level of zero-energy Raman shift. The integration time was set to 130 ms.

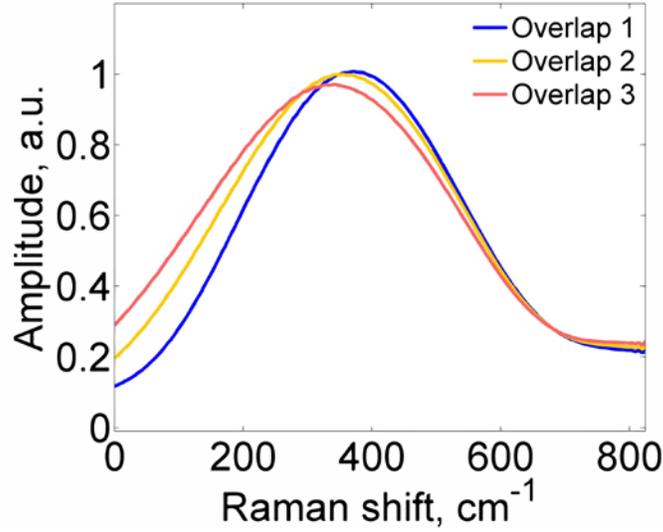

FIG. S8. 1D detection windows in the cases of *overlap1* (blue), *overlap2* (yellow) and *overlap3* (light red). The shapes were obtained by integrating corresponding 2D detection windows along the horizontal (*detection energy*) axis.

We notice that Raman features, which are in the close proximity to the zero-frequency feature, are obscured by a large background from the non-oscillating contributions, whereas others are enhanced (Fig. S9).

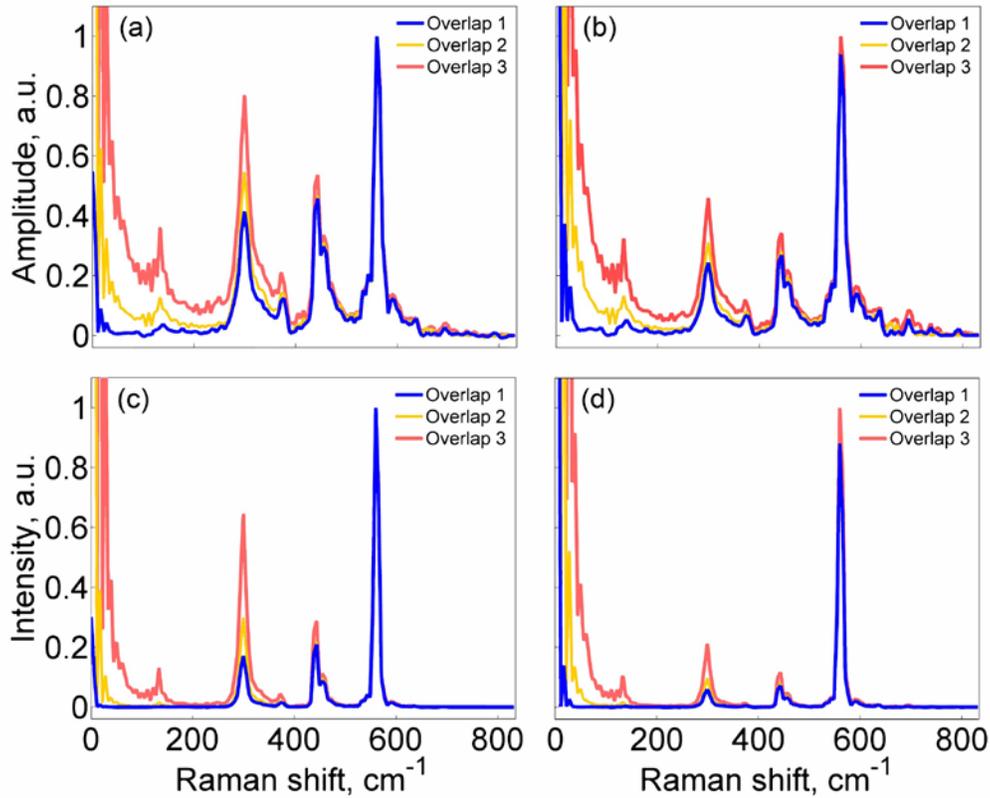

FIG. S9. (a) 1D Raman (CSRS) amplitude spectra in the cases of *overlap1* (blue), *overlap2* (yellow) and *overlap3* (light red) between the spectral shapes of the pump and the Stokes pulses, obtained by integrating corresponding 2D spectral maps along the horizontal (*detection energy*) axis, and (b) corresponding calibrated 1D Raman amplitude spectra. (c,d) 1D Raman intensity spectra, obtained by squaring 1D Raman amplitude spectra in (a) and (b), respectively.

## 6. Raman spectra of IR-813, IR-806 and IR-140

2D and integrated CSRS spectra of the NIR cyanine dyes IR-813, IR-806 and IR-140, shown in Fig. S10, were acquired with the pulses, spectral shapes of which are shown in Fig. S7a.

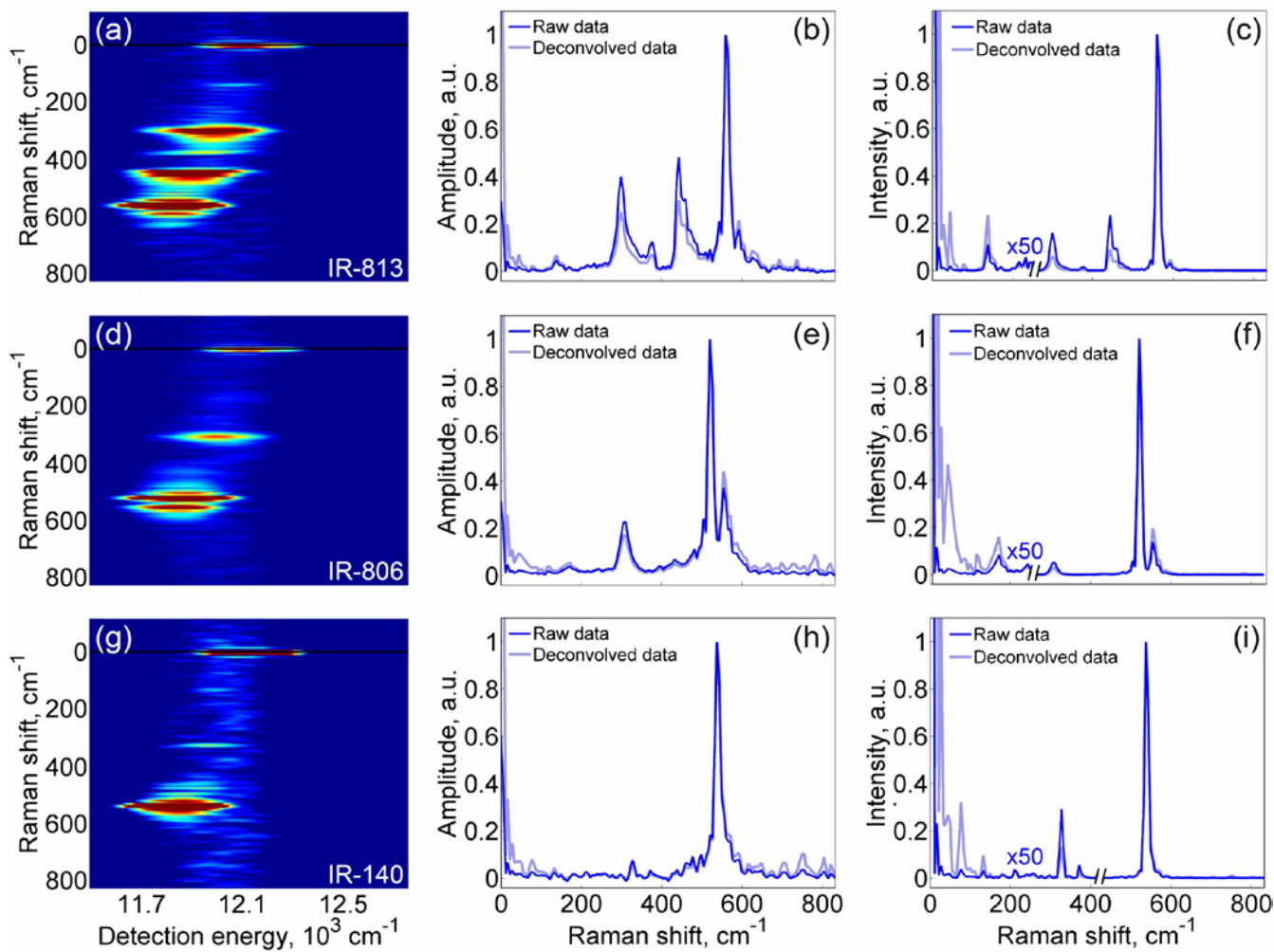

FIG. S10. (Left column) measured 2D CSRS maps of electric field amplitude as a function of the detection energy and the observed Raman shifts for (a) IR-813, (d) IR-806 and (g) IR-140; (middle column) integrated along the detection energy axis 1D amplitude Raman spectra (blue) and corresponding calibrated data (light blue) for (b) IR-813, (e) IR-806 and (h) IR-140; (right column: c,f,i) 1D intensity Raman spectra and corresponding calibrated data, derived from the 1D amplitude Raman spectra by squaring. The intensities in the lower wavenumber region are scaled by a factor of 50. The integration time was set to 130 ms.

The Raman spectra for the IR-806, to the best of our knowledge, are not reported in the literature. We observe low-energy low-intensity Raman vibrations below 200 cm$^{-1}$ of the dyes, which, to our knowledge, have also never been reported previously. The frequencies of all Raman modes found for the IR-813, IR-806 and IR-140 are summarized in the Table 2.

TABLE 2. Measured Raman modes of the NIR cyanine dyes IR-813, IR-806 and IR-140.

| Sample | Raman shifts, ±1.3 cm$^{-1}$ |
| --- | --- |
| IR-813 | 33.2 cm$^{-1}$, 77.6 cm$^{-1}$, 138.5 cm$^{-1}$, 299.2 cm$^{-1}$, 376.8 cm$^{-1}$, 443.3 cm$^{-1}$, 559.7 cm$^{-1}$, 599 cm$^{-1}$, 626.2 cm$^{-1}$, 642.8 cm$^{-1}$, 698.2 cm$^{-1}$, 731.5 cm$^{-1}$, 737 cm$^{-1}$, 770.3 cm$^{-1}$, 814.6 cm$^{-1}$ |
| IR-806 | 44.3 cm$^{-1}$(?)[a], 116.4 cm$^{-1}$, 171.8 cm$^{-1}$, 310.3 cm$^{-1}$, 393.4 cm$^{-1}$, 432.2 cm$^{-1}$, 520.9 cm$^{-1}$, 554.2 cm$^{-1}$, 665 cm$^{-1}$(?)[a], 698.2 cm$^{-1}$(?)[a], 725.9 cm$^{-1}$(?)[a], 781.4 cm$^{-1}$ |
| IR-140 | 77.6 cm$^{-1}$(?)[a], 133 cm$^{-1}$, 182.9 cm$^{-1}$(?)[a], 210.6 cm$^{-1}$, 326.9 cm$^{-1}$, 371.3 cm$^{-1}$, 537.5 cm$^{-1}$, 592.9 cm$^{-1}$, 648.4 cm$^{-1}$, 703.8 cm$^{-1}$(?)[a], 753.6 cm$^{-1}$(?)[a], 803.5 cm$^{-1}$(?)[a] |

[a]These Raman modes were detected either in the region of a significant contribution from zero-energy peak or with low S/N ratio.

## 7. Spontaneous Raman measurements

We measured spontaneous Raman spectra of the IR-806 and IR-813 using conventional Raman spectroscopy technique (Renishaw confocal Raman microscope). Fig. S12a shows the measurements for the IR-806, when the excitation wavelength (785 nm) was chosen to be close to the wavelength we used for our CSRS measurements. In this case Raman modes are greatly obscured by the strong fluorescence background, and no Raman features can be discerned.

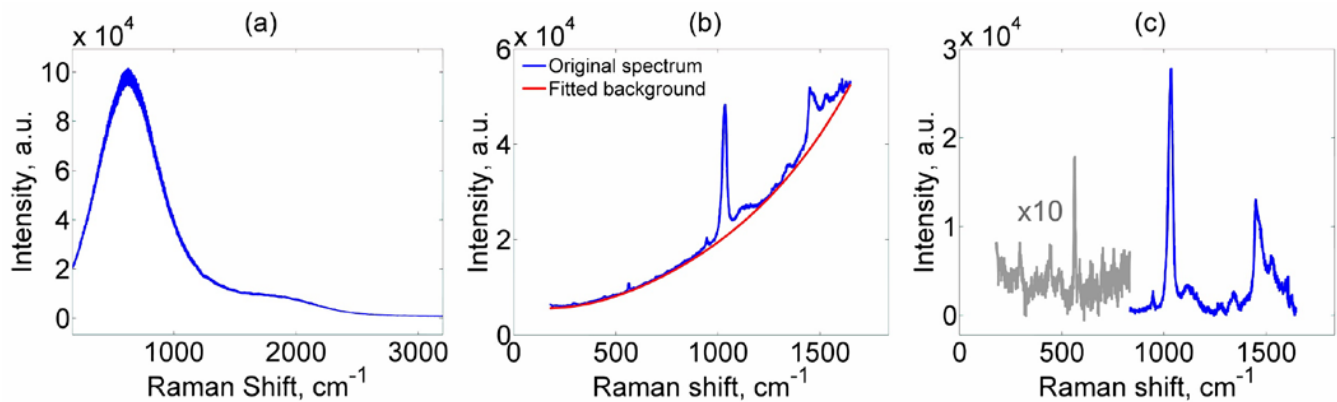

FIG. S12. (a) Fluorescence background, completely obscuring Raman modes of the IR-806. The measurements were done at the excitation wavelength of ~785 nm, with a 20x objective lens. The integration time was 10 sec. (b) Spontaneous Raman spectra of the IR-813 measured with a 50x objective lens at the excitation wavelength ~488 nm. Integration time was 50 sec and the volume probed was ~0.4 pL. The fitted background[8] is represented by the red line. (c) Spontaneous Raman spectra of the IR-813 after the background subtraction. The part of the spectrum, scaled by a factor of 10 and matching our detection range is colored in grey.

Fig. S12b,c represent the spontaneous Raman spectra, measured with the excitation wavelength of 488 nm (to minimize the effects of fluorescent background) and integration time 50 sec (to be able to detect low-energy vibrational modes). In this case, the tail of the fluorescent background is still present, however, due to its reduced contribution, it is possible to discern Raman features. We applied background subtraction technique, described in,[8] to obtain background-free Raman spectra and compared them with the corresponding CSRS spectra (Fig. 5c of the main manuscript).

## 8. Anti-Stokes-to-Stokes ratio in the limit of low energy vibrations

The effective anti-Stokes-to-Stokes ratio $\rho_{\text{eff.}}$ can be represented by the following expression,[9]:

$$\rho_{\text{eff.}} = \rho \frac{\sigma_{\text{AS}}}{\sigma_{\text{S}}} = \frac{I_{\text{AS}}}{I_{\text{S}}} = \left[\left(\frac{\upsilon_{\text{laser}} + \delta\upsilon_{\text{vib.}}}{\upsilon_{\text{laser}} - \delta\upsilon_{\text{vib.}}}\right) e^{-\frac{h\delta\upsilon_{\text{vib.}}}{kT}}\right] \frac{\sigma_{\text{AS}}}{\sigma_{\text{S}}}, \qquad (9)$$

where $\rho$ is the "intrinsic" Stokes-to-anti-Stokes ratio, when the excitation frequency is tuned far from any electronic resonances; $I_{\text{AS}}$ and $I_{\text{S}}$ are, correspondingly, the anti-Stokes and Stokes Raman intensities, $\upsilon_{\text{laser}}$ is the frequency of the excitation photon, $\delta\upsilon_{\text{vib.}}$ is the Raman shift of a vibrational mode, $h$ is the Planck's constant, $T$ is the temperature, $k$ is the Boltzmann constant. Here, the multipliers $\sigma_{\text{AS}}$ and $\sigma_{\text{S}}$ take into account various modulation effects that could affect the ratio $\rho$, such as the resonant enhancement (which in some cases can lead to the anti-Stokes features being more intense than Stokes peaks), the finite spectral width in the case of the pulsed excitations (leading to the modulation by cross-correlation window function), and others. In the case of low-energy detection limit $|\delta\upsilon_{\text{vib.}}|\ll 1$, the first-order Taylor expansion of (9) leads to the following expression:

$$\rho_{\text{eff.}} \approx \left[1 + \left(\frac{8}{\upsilon_{\text{laser}}} - \frac{h}{kT}\right)\delta\upsilon_{\text{vib.}}\right]\frac{\sigma_{\text{AS}}}{\sigma_{\text{S}}} = \left(1 + \tilde{c}\delta\upsilon_{\text{vib.}}\right)\frac{\sigma_{\text{AS}}}{\sigma_{\text{S}}}, \qquad (10)$$

where the constant $\tilde{c} < 0$ for all frequencies in the visible and NIR range $\upsilon_{\text{laser}}$, rendering the intensities of the low-energy anti-Stokes Raman peaks being intrinsically always lower than the intensities of the corresponding Stokes Raman features, provided the excitation light is tuned far from the electronic resonances. However, in reality it is not always possible to perform the measurements, when $\sigma_{\text{AS}} = \sigma_{\text{S}} = 1$, and frequently it is desirable to have resonant enhancement, which is usually different on the Stokes and the anti-Stokes detection sides. This provides another justification for having the ability to easily switch between the CSRS- and CARS-type measurements: depending on the specific experimental conditions, it is often preferable to perform the measurements in a more sensitive detection region.